\documentclass[12pt]{iopart}
\pdfoutput=1
\usepackage{hyperref}
\usepackage{enumerate}
\usepackage{setspace}
\usepackage[square,sort,comma]{natbib}

\usepackage{graphicx}

\newcommand*{\req}[1]{(\ref{#1})}

\begin{document}

\title{Distributed authentication for randomly compromised networks}

\author{Travis R. Beals$^1$, Kevin P. Hynes$^2$, and Barry C. Sanders$^2$}

\address{$^1$ Department of Physics, University of California, Berkeley, California 94720, USA}
\address{$^2$ Institute for Quantum Information Science, University of Calgary, Alberta T2N 1N4, Canada} 
\ead{bsanders@qis.ucalgary.ca}

\date{}

%\keywords{quantum key distribution; secret sharing; information theoretic security; authentication}

\begin{abstract}
We introduce a simple, practical approach with probabilistic in\-for\-ma\-tion-theoretic security to solve one of quantum key distribution's
major security weaknesses: the requirement of an authenticated classical channel to prevent man-in-the-middle attacks.
Our scheme employs classical secret sharing and partially trusted intermediaries to provide arbitrarily high confidence in the security of the protocol. 
Although certain failures elude detection, we discuss preemptive strategies to reduce the probability of failure
to an arbitrarily small level: probability of such failures is exponentially suppressed with increases in connectivity (i.e., connections per node).
\end{abstract}

\pacs{03.67.Dd, 89.20.Ff, 89.70.-a, 89.75.-k}

%\submitto{\NJP}

\maketitle

\section{Introduction\label{sec:introduction}}

The security of all public key cryptosystems depends on assumptions about the hardness of certain mathematical problems. These assumptions are unproven, making public key cryptography vulnerable to advances in cryptanalysis.  In 1994, Peter Shor made such an advance by developing an efficient algorithm for prime factorization and discrete logarithms~\cite{shor:factor2}. Shor's algorithm breaks the most common public key cryptosystems but requires a moderately powerful quantum computer, which fortunately does not yet exist. 

Given the importance of public key cryptography to the internet and electronic commerce, it is desirable that a practical information-theoretic secure replacement be developed and implemented well before public key cryptosystems become vulnerable to quantum computers or other attacks ~\cite{paterson:whyqc,stebila:qkd}. Quantum key distribution (QKD), first developed by Bennett and Brassard~\cite{bennett:BB84}, provides a partial solution when coupled with one-time pad (OTP) encryption. QKD relies on the uncertainty principle of quantum mechanics to provide information-theoretic security against eavesdropping, but unfortunately QKD requires authenticated classical channels to prevent man-in-the-middle (MITM) attacks. Information-theoretic secure protocols exist for the authentication of classical channels~\cite{wegman:universal,wegman:newhash,stinson:universal}, but such protocols require a shared secret key; this requirement is difficult for mutual strangers to satisfy. We refer to the requirement of authenticated channels as the \emph{stranger authentication problem}; QKD must overcome this problem to become a feasible, secure alternative to today's cryptographically-secure public key systems.

We propose a solution to the stranger authentication problem by encoding an authentication key into multiple shares~\cite{shamir:secret,blakley:313}. These shares are transmitted via multiple paths through a QKD network in which some nodes already share secret keys. Our approach prevents MITM attacks with high probability, even if the attacker controls a large, randomly-selected subset of all the nodes.
As authenticated QKD in combination with OTP provides information-theoretic security~\cite{renner:012332}, we describe the level of security of our protocol as \emph{probabilistic information-theoretic}. By this, we mean that the security of our protocol is stochastic; with very high probability, the protocol provides information-theoretic security and with some very small probability $\delta$, it fails in such a way as to allow a sufficiently powerful adversary to perform undetected MITM attacks. The security parameter $\delta$ can be made arbitrarily small by modest increases in resource usage.

\section{Adversary and Network Model}

It is convenient to model networks with properties similar to those described above by using undirected graphs, where each vertex represents a node or party participating in the network and each edge represents an authenticated public channel. Such a channel could be provided by using a shared secret key for authentication, or by any other means providing information-theoretic security. We also assume that all parties in the network are connected to all other parties by unauthenticated channels that allow both classical and quantum information. 

This last point can be restated as the assumption that the network of (unauthenticated) quantum channels is the complete graph. In the case of a geographically large QKD network using present-day technology, distance limitations of point-to-point QKD links would make this assumption challenging to satisfy, although protocols have been proposed to address this problem under certain circumstances\cite{beals:distrelay,salvail:repeater}. In the long term, we believe quantum repeaters~\cite{duan:6862} will overcome QKD's distance limitations.

\subsection{Adversarial capabilities\label{sec:adv_model}}
We consider an adversary, which we will call the \emph{sneaky supercomputer}~\cite{beals:distrelay}:
\begin{enumerate}[(i)]
\item The adversary is computationally unbounded.
\item The adversary can listen to, intercept, and alter any message on any public channel.
\item The adversary can compromise a randomly-selected subset of the nodes in the network. Compromised nodes are assumed to be under the complete control of the adversary. The total fraction of compromised nodes is limited to $(1-t)$ or less.
\end{enumerate}
This adversary would be at least as powerful as one with a quantum computer. It can successfully perform MITM attacks against public key cryptosystems (using the first capability) and against unauthenticated QKD (using the second capability) but not against a QKD link between two uncompromised nodes that share a secret key for authentication (since quantum mechanics allows the eavesdropping to be detected)~\cite{renner:012332}. The adversary can always perform denial-of-service (DOS) attacks by simply destroying all transmitted information; since DOS attacks cannot be prevented in this adversarial scenario, we concern ourselves only with security against MITM attacks and do not consider robustness against DOS attacks further.

The third capability in this adversarial model,
namely the adversary's control of a random subset of nodes,
simulates a network in which exploitable vulnerabilities are present on some nodes but not others.
As a first approximation for a real-world network, we assume that vulnerable nodes are randomly distributed throughout the network. Others have considered adversarial models in which the adversary cannot compromise nodes~\cite{elliott:quantum,secoqc:whitepaper}, or can deterministically compromise a small number of nodes~\cite{salvail:repeater}.

\subsection{The Network}
For the stranger authentication problem, let us represent the network as a graph~$G$, with~$V(G)$ being the set of vertices (nodes participating in the network) and $E(G)$ being the set of edges (secure authenticated channels, e.g.~QKD links between parties who share secret keys for authentication). 
We denote $N = |V(G)|$ as the number of vertices (nodes). 
The set of compromised nodes~$V_d$ is assumed to be controlled by the adversary: $|V_d| \leq N (1-t)$.

In Section \ref{sec:authentication}, we describe a protocol that allows an arbitrary uncompromised Alice and Bob ($A, B \in V(G)\backslash V_d$) who do not initially share a direct link (i.e., $(A,B) \notin E(G)$) to communicate with information-theoretic security with very high probability. 
Specifically, we show that Alice and Bob can generate a shared secret key via secret sharing
with shares transmitted via multiple paths through the graph.
Moreover Eve's probability of learning the secret shared key is 
smaller than~$\delta$, which Alice and Bob can make arbitrarily small. 
Once Alice and Bob share a secret key, they use it to authenticate a QKD link with each other.
In other words, we assume that Alice and Bob can easily acquire an unauthenticated QKD link 
then use their shared secret key to authenticate that link.

With reference to the adversary's third capability, we can consider an alternate scenario in which  the adversary is instead able to choose a subset of nodes to compromise. If the adversary can control at most $n$ nodes, then $n+1$ node-disjoint paths between Alice and Bob are required to guarantee security~\cite{salvail:repeater}.

\subsection{Other approaches}

We could model existing public key-based networks using a similar convention, where edges represent authenticated (but not necessarily secure) channels used for initial distribution of public keys. Such graphs typically have a tree topology, in which most parties are connected only to a single root certificate authority (CA), or small number of root certificate authorities. Before participating in a public key network, users typically obtain a copy of a certificate (which contains the CA's public key) from a few certificate authorities. This certificate is then used to verify the authenticity of digitally-signed public keys presented by other parties, thereby thwarting MITM attacks. 

The process of obtaining the CA's certificate is often hidden from the user because the certificate is usually bundled with their web browser or operating system. Nonetheless, obtaining an authentic root CA certificate is a crucial part of the user's security process, as a compromised CA certificate would allow an attacker to subvert all future communication performed by that user.

We are now able to make a few observations about the existing public key infrastructure. First of all, it relies critically on the honesty and reliability of a small number of root CAs. While the root CAs have so far proved to be honest, they do occasionally make mistakes. Second, public key networks of $N$ parties require only $O(N)$ authenticated channels to perform the initial distribution of root CA certificates. Third, the compromise of a single authenticated channel can result in compromise of all future communication involving a particular single user, but the rest of the parties are largely protected. Finally, as is mentioned earlier, public key-based systems are vulnerable to our ``sneaky supercomputer'' adversary (e.g., an attacker with a quantum computer) and the consequences of a successful attack against any of the root CAs would be severe.

To solve the stranger authentication problem, a na\"{i}ve approach is represented by a complete graph wherein every vertex is connected to every other vertex. 
This approach has the drawback that each party would have to have a unique secret key shared with each other party, implying $O(N^2)$ secret keys in total. Alternatively, we could seek to duplicate the tree topology of public key networks and have a few central trusted parties to which all others are connected~\cite{pasquinucci:0506003}. This has most of the drawbacks listed above for public key networks, as well as the additional requirement that the central trusted parties be actively involved in every communication (rather than one-time signing of public keys). Clearly, both of these na\"{i}ve approaches suffer from serious limitations that make them impractical for large-scale networks. In the subsequent section, we present efficient protocols for solving these problems.

\section{The Stranger Authentication Protocol\label{sec:authentication}}

In this section, we describe the stranger authentication protocol in the context of a uniform random graph. We analyze its security and efficiency in the context of this network topology, and in  Section \ref{sec:topologies}, we introduce the power-law topology, which we study via numerical simulations.

Consider a uniform random graph $G$ with $N$ vertices, in which the $|E(G)|$ edges are random, in the sense that each possible edge $e \in V^2(G)=V(G)\times V(G)$ is equally likely to be a member of the set of edges, $E(G)$.  Furthermore, we assume that the set of compromised nodes $V_d$ is a randomly-chosen subset of the total set of vertices, $V(G)$. Take some arbitrary uncompromised Alice and Bob in $G$: if Alice and Bob are acquaintances and share a secret key (i.e., $(A,B) \in E(G)$), then they can communicate securely using QKD to generate a large key for use as a OTP, using their shared secret key to authenticate their QKD exchange. We are concerned with the case in which Alice and Bob are mutual strangers (i.e., $(A,B) \notin E(G)$).

\begin{figure}
 \includegraphics[width=400px]{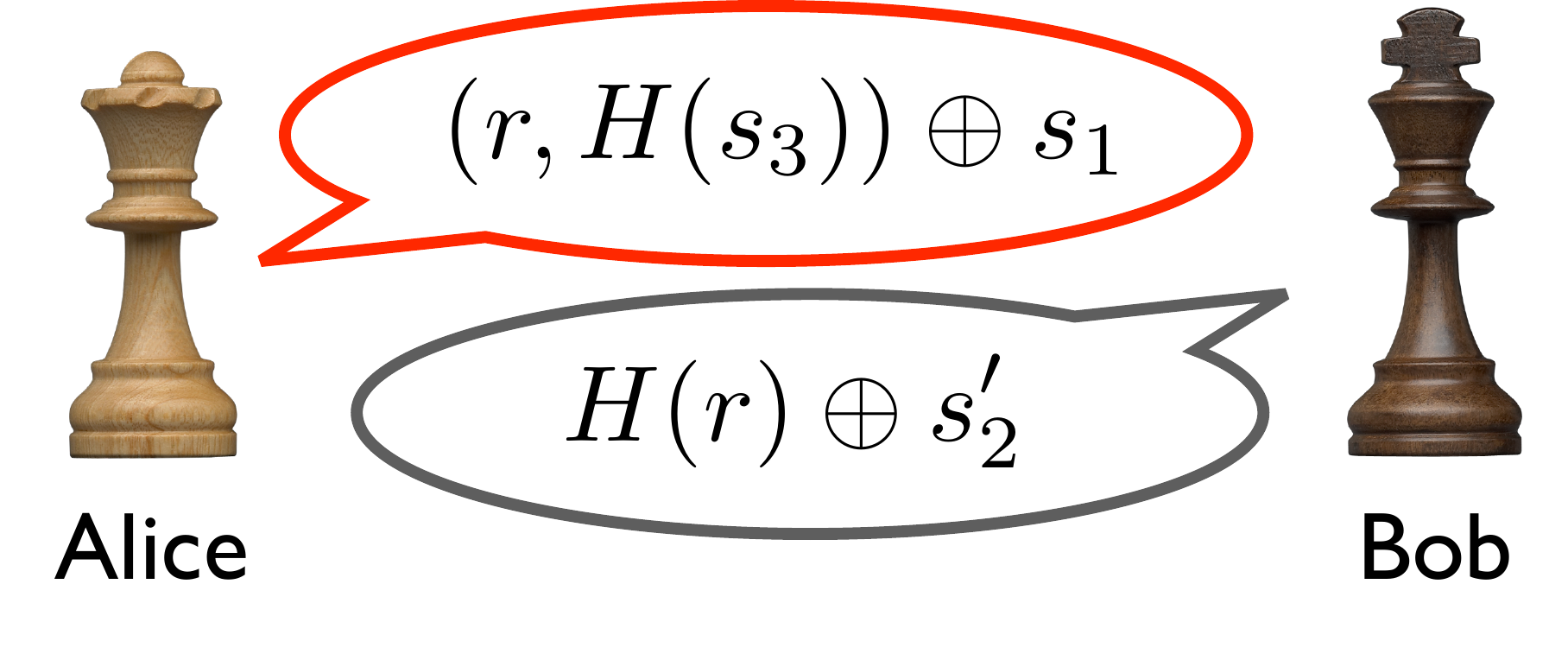}
 \caption{\label{fig:protocol} Alice and Bob verify that their respective secret keys,~$s=(s_1,s_2,s_3)$ and~$s^\prime=(s_1^\prime,s_2^\prime,s_3^\prime)$, are in fact the same through the exchange shown above.  Alice generates a random number $r$, concatenates it with the hash $H(s_3)$ of~$s_3$, XORs this with~$s_1$ and sends the result to Bob. Bob decodes with~$s_1^\prime$, verifies that $H(s_3) = H(s_3^\prime)$, then sends back to Alice the result of bit-wise XORing the hash of $r$, $H(r)$, with~$s_2^\prime$. Finally, Alice decodes with~$s_2$ and checks to see that the value Bob has computed for $H(r)$ is correct. Alice and Bob now know~$s_3 = s_3^\prime$ and can store~$s_3$ for future use. The lengths of~$s_1$ and~$s_2$ scale as $O(-\log \delta)$, where $\delta$ is the maximum allowable probability that an attacker who does not know~$s$ can modify~$s^\prime$ and escape detection.   The length of~$s_3$ is therefore only slightly less than $l$ (the length of~$s$). Alice and Bob thus choose $l$ so that the length of~$s_3$ will be sufficient for their purposes. Note that with this protocol, Eve can fool Alice and Bob into accepting~$s \neq s^\prime$ with 100 \% probability if Eve knows~$s$ and~$s^\prime$. }
 \end{figure}

In order for Alice to communicate securely with Bob, she first needs to establish a small shared secret key with him to prevent MITM attacks. To obtain a shared secret key, Alice and Bob use the following procedure:
\begin{enumerate}[(i)]
  \item Alice generates a random string of length $l$:~$s \in \{0,1\}^l$. $l$ is chosen as described in Fig.~\ref{fig:protocol}.
  \item Alice selects all cycle-free paths between her and Bob (we assume Alice and Bob have a complete and accurate routing table for the graph). Define $n$ to be the number of such paths.
  \item Alice employs a secret sharing scheme~~\cite{shamir:secret,blakley:313} to encode~$s$ into $n$ shares, such that all~$n$ shares are required to reconstruct~$s$.
  \item Alice sends Bob one share via each cycle-free path.
  \item Bob receives the shares and combines them to obtain~$s^\prime$.
  \item Alice and Bob use the protocol described in Fig.~\ref{fig:protocol} to determine if~$s=s^\prime$. If so, they are left with a portion of~$s$ (identified as~$s_3$), which is their shared secret key. If~$s \neq s^\prime$, Alice and Bob discard~$s$ and~$s^\prime$ and repeat the protocol.
\end{enumerate}

The protocol as described above is not efficient, in that the number $n$ of cycle-free paths  grows rapidly as the size of the graph increases. The protocol can be made efficient without much loss in security, as we show later.

There are some similarities between our protocol and that described by Dolev et al.~\cite{dolev:perfect} for networks of processors with some faulty processors, in that both protocols use secret sharing as a means of protecting against nodes that deviate from the protocol.
The protocols differ substantially with respect to the adversarial models they address.
Dolev et al.'s protocol handles cases wherein the adversary controls a small number of nodes of his or her choice.
In contrast our protocol is designed for a network in which the adversary controls a large fraction of all the nodes but cannot choose which nodes are compromised.

\subsection{Security}
Provided there exists at least one path between Alice and Bob with no faulty or uncompromised nodes,
Alice and Bob are authenticated and acquire a shared secret key. 
MITM attacks are prevented because each node in the path authenticates the nodes before and after, thereby providing a chain of authentication.

In order for an attacker (Eve) to compromise the protocol, the attacker must learn~$s$.
As~$s$ is encoded into multiple shares, and all shares are required to reconstruct~$s$,
the attacker must learn all the shares in order to learn~$s$. Shares are secured in transmission by OTP using keys generated by authenticated QKD, so the only way for Eve to learn a share is if she controls a node that is part of the transmission path between Alice and Bob for that share. Put another way, the protocol is secure unless all share transmission paths between Alice and Bob contain at least one compromised node. 

\begin{figure}[p] \centering
\includegraphics[width=2.50 in,  keepaspectratio=true]{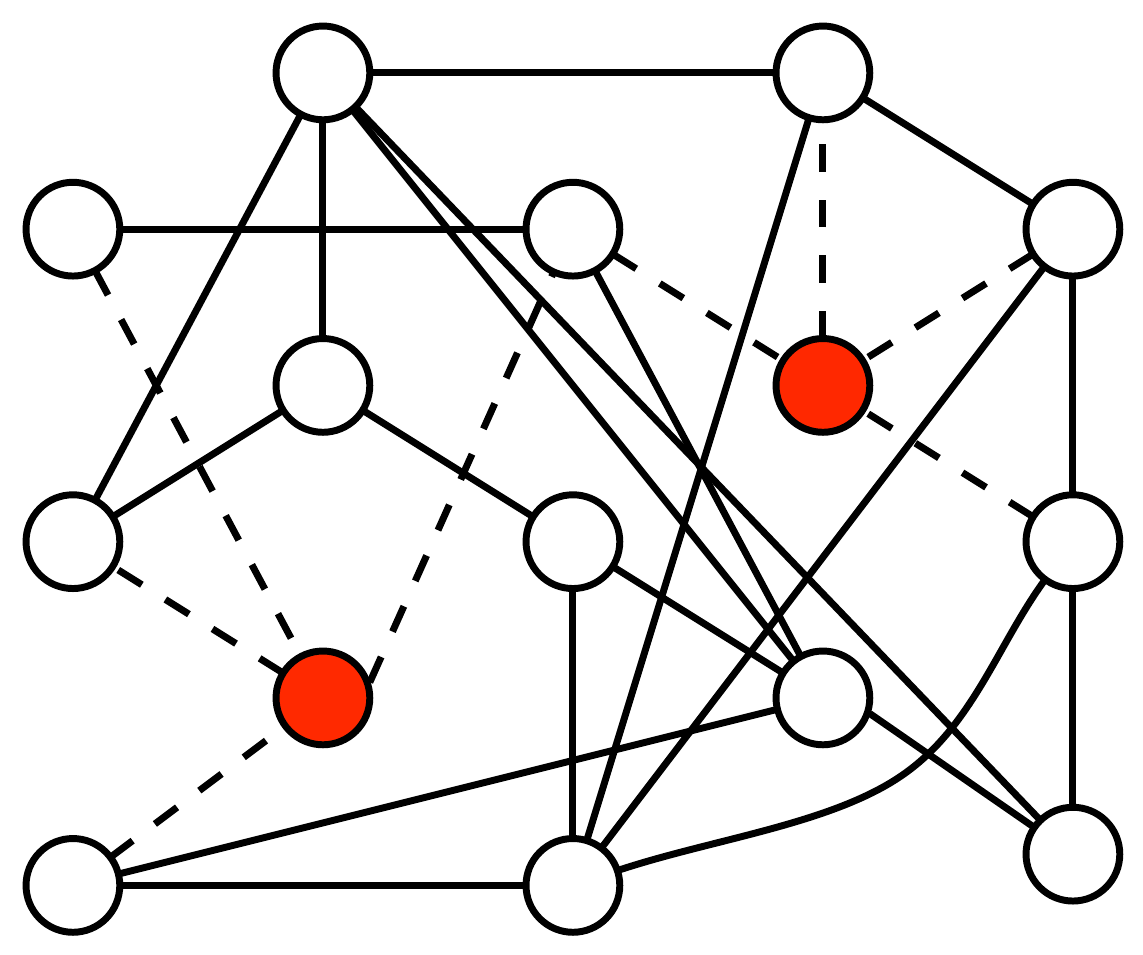}
\caption{\label{fig:graph} White vertices represent honest parties, whereas shaded vertices represent dishonest parties. Dashed edges are those that end on a dishonest party. The subgraph induced by the removal of the shaded vertices remains connected, so any two honest parties can communicate securely using the protocol described in Sec. \ref{sec:authentication}.}
\end{figure}

This question can be rephrased in graph-theoretic terms by asking how large $E(G)$ must be such that the subgraph $G^\prime \subseteq G$ induced by $V(G)\backslash V_d$ is connected. As an example, Fig.~\ref{fig:graph} shows a random graph that remains connected even after two compromised parties are removed.

Recall that $$t \equiv 1- \frac{|V_d|}{|V(G)|}$$ is the fraction of uncompromised nodes.
Thus, $\lfloor t N\rfloor$ nodes and approximately $t^2 |E(G)|$ edges will remain after the compromised nodes are removed. Using a result obtained by Erd\H{o}s and R\'enyi~\cite{erdos:randomgraphsI} for connectedness of uniform random graphs, we see that, for a random graph of $\lfloor t N\rfloor$ vertices and $\lfloor t^2 |E(G)|\rfloor$ edges in the limit as $t N \rightarrow \infty$, the probability that the graph is connected is
\begin{equation}
	p_c = \rme^{-\rme^{-2c}}
\end{equation}
with
\begin{equation}
	c = \frac{\left|E(G) \right| t}{N} - \frac{1}{2} \log{t N}.
\end{equation}

Suppose we wish to estimate the number of edges, $|E(G)|$, required to limit the probability of compromise to some $\epsilon = 1 - p_c$. Using the above result for $$1 \gg \delta \gg t N \exp \left[-t (N-1) \right],$$
we obtain the following relation:
\begin{equation}
	|E(G)|  \simeq  \frac{1}{2} \frac{N}{t} \log \left( \frac{N t}{\epsilon} \right). 
\label{eq:edges}
\end{equation}

This estimate does not apply in the case where N is finite and $\epsilon = 0$, where it is clear the number of edges goes to $O(N^2)$ rather than infinity. It is worth noting that Eq.~\req{eq:edges} gives the approximate number of edges required such that, with probability $1-\epsilon $, \emph{all} $A$, $B$, in $V(G) \backslash V_d$ can communicate securely. This is a stronger result than if we had shown secure communication just for one particular choice of $A$ and $B$.

\subsection{Efficiency and Cost}\label{sec:efficandcost}

According to the protocol described above, the number of shares $n$ required to perform the protocol between two arbitrary parties (Alice and Bob) will grow with the total number of cycle-free paths between them and thus much faster than the total number of parties. This is not likely to be a serious problem for small-to-medium networks as the strings being sent need not be large (they are necessary only for the initial authentication of a QKD protocol) and need only be sent the first time a given pair of parties wish to communicate. Nonetheless, it will eventually become a problem as the network grows larger so we seek to reduce this cost by having Alice and Bob use only a subset of the possible paths. Fortunately, relatively few paths are required to guarantee a high degree of security, and the number of paths required scales slowly with the size of the graph.

We can estimate the number of paths required to keep the probability of compromise $\delta$ constant. (Note that the following calculation is intended as an estimate rather than a rigorous derivation.)
The probability of a single path of length $\ell$ being uncompromised is approximately $t^{\ell - 1}$. Suppose we have a graph $G$ with the number of edges $|E(G)|$ chosen according to Eq.~\req{eq:edges}, where $\epsilon \leq \delta$.

The diameter (maximal distance between any pair of nodes) of such a graph~\cite{albert:randomgraph} is
\begin{equation}
	d\simeq\frac{\log N}{\log{\left\langle k \right \rangle}} \label{eq:d}
\end{equation}
with
\begin{equation}
	\left\langle k \right \rangle=2 |E(G)| /N
\end{equation}
the average degree of a node in $G$. 
Expression~\req{eq:d} can be understood by considering the number of nodes at distance $d$ from some starting node; since each node has on average $\left\langle k \right \rangle$ neighbours, we expect to find  $\left\langle k \right \rangle^d$ nodes at this distance.

If we wish to find the $p$ shortest (not necessarily independent) paths from Alice to Bob, we can estimate the length of the longest of these paths as $$\ell_p =  \frac{\log (N p)}{\log{\left\langle k \right \rangle }}.$$
We can now estimate the number $p$ of independent paths required to give a probability $\delta$ that all paths are compromised:
\begin{eqnarray}
	p  =  \frac{\log \delta}{\log \left( 1-t^{\ell_p - 1} \right)}.
\label{eq:preq}
\end{eqnarray}
Note that the right-hand side of Eq.~\req{eq:preq} contains $\ell_p$, which depends on $p$; an iterative process can easily be used to find a solution for $p$.

We have estimated the number of independent paths~\req{eq:preq} required to guarantee a certain degree of security; however, it will in general be more practical to choose paths that are not completely independent of each other. This will result in some reduction of security, which can be compensated for by using additional paths. Another issue is that path choice must not be determined solely by a dishonest party; if path choice is left solely to Alice, and a malicious Eve wants to impersonate Alice, Eve can choose paths such that every path contains a compromised node. One solution is for Bob to initiate a second round of the protocol (using different paths) to confirm that he is indeed communicating with Alice; the shared secret keys generated in the two rounds are then combined to produce a final key.

We use Monte Carlo simulations to test whether our estimate of path requirements is reasonable and to compare the performance of uniform random graphs with that of power-law graphs. The path-picking algorithm used in this simulation is based upon Dijkstra's algorithm~\cite{cormen:introalgorithms} for solving the shortest path problem. Initially, all nodes are given the same cost. The path of least cost (``shortest'' path) between Alice and Bob is chosen, and the cost of all intermediate nodes along the path are incremented. A new path of least cost is found, and the process is repeated until the desired number of paths are chosen. Duplicate paths are not allowed.

In a real-world application, a higher number of paths used would likely be chosen so as to provide a much lower risk of compromise (smaller $\delta$); parameter values used in this simulation are for illustrative purposes. We make no claims as to the optimality of the path-picking algorithm used, but note that it gives a nearly constant degree of security (i.e., $\delta$ is nearly constant) with resource use consistent with the modest scaling with $N$ predicted by our estimates.  Simulation details and results are given in the following two subsections.

\begin{figure*}[p] \centering
\includegraphics[width=2.5 in,  keepaspectratio=true]{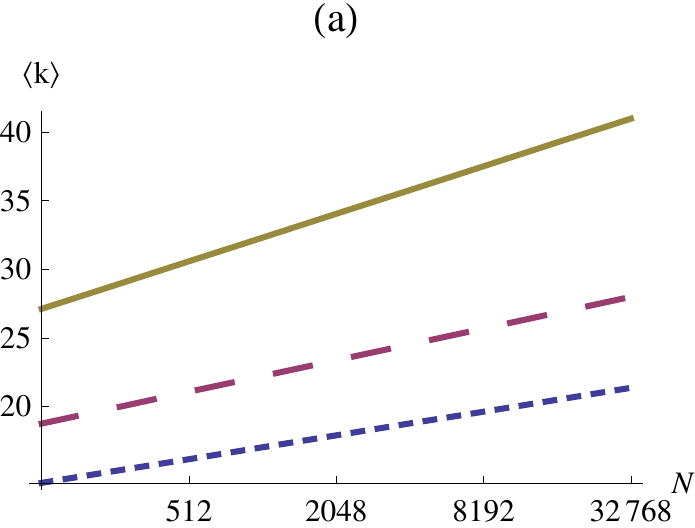}
\includegraphics[width=2.5 in,  keepaspectratio=true]{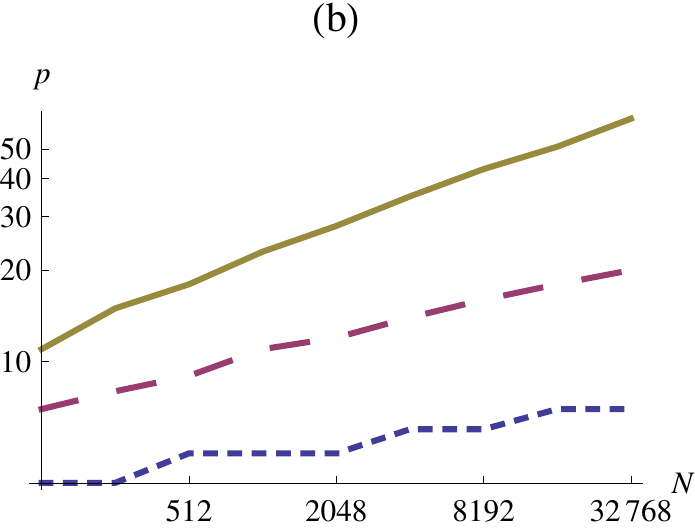}
\caption{\label{fig:parameters}Parameters used in generating graphs for all simulations, both for the uniform random and truncated power-law cases. The yellow solid line, red long-dashed line, and blue short-dashed line represent values for $t=0.4$, $t=0.6$, and $t=0.8$, respectively. (a) shows the average number of edges per node, $\langle k \rangle /2$, as a function of $N$, the total number of nodes, and (b) indicates the number of paths p used as a function of the number of nodes, $N$. 
}
\end{figure*}

\begin{figure*}[p] \centering
\includegraphics[width=3 in,  keepaspectratio=true]{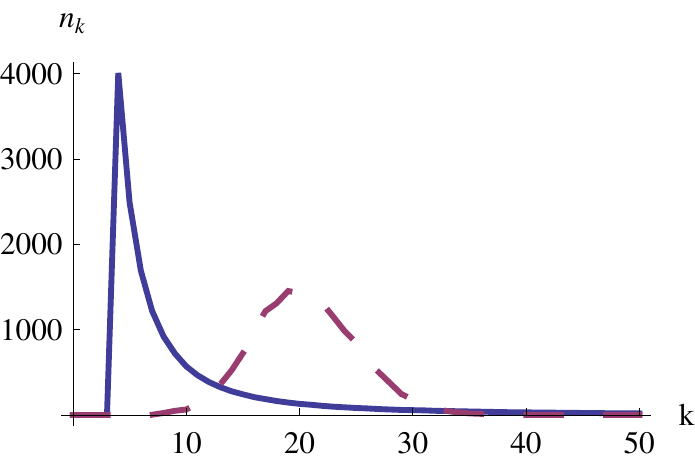}
\caption{\label{fig:degreecomparison}A comparison of the degree distribution (number of nodes $n_k$ with $k$ edges) of the two graph types being considered with two sample graphs for which $N=16384$, $t=0.8$, and $|E(G)| = 167821$. The red dashed line shows the uniform random case while the solid blue line is the truncated power-law case.
}
\end{figure*}

\begin{table*}[p] \centering
\begin{tabular}{|c|c c c|c c c|}
\hline\hline
Nodes & \multicolumn{3}{c|}{Uniform Random} & \multicolumn{3}{c|}{Power-Law} \\
 & $t=0.8$ & $t=0.6$ & $t=0.4$ & $t=0.8$ & $t=0.6$ & $t=0.4$ \\
\hline
128   & 3 & 3 & 3 & 4 & 4 & 3 \\
256   & 4 & 4 & 3 & 4 & 4 & 4 \\
512   & 4 & 4 & 4 & 4 & 4 & 4 \\
1024  & 4 & 4 & 4 & 4 & 4 & 4 \\
2048  & 4 & 4 & 4 & 4 & 4 & 4 \\
4096  & 5 & 5 & 5 & 4 & 4 & 4 \\
8192  & 5 & 5 & 5 & 4 & 4 & 4 \\
16384 & 5 & 5 & 5 & 4 & 4 & 4 \\
32768 & 5 & 5 & 5 & 4 & 4 & 5 \\
\hline
\end{tabular}
\caption{\label{table:longestpath}The length of the $99^{\mathrm{th}}$-percentile longest path.}
\end{table*}

\begin{figure*}[p] \centering
\includegraphics[width=2.5 in,  keepaspectratio=true]{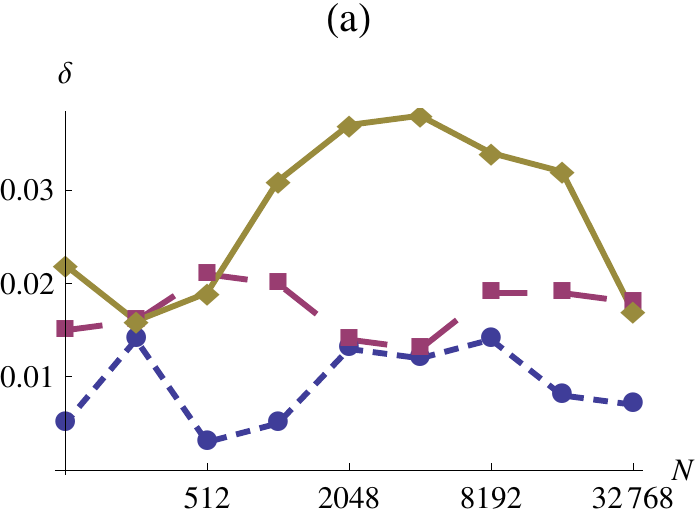}
\includegraphics[width=2.5 in,  keepaspectratio=true]{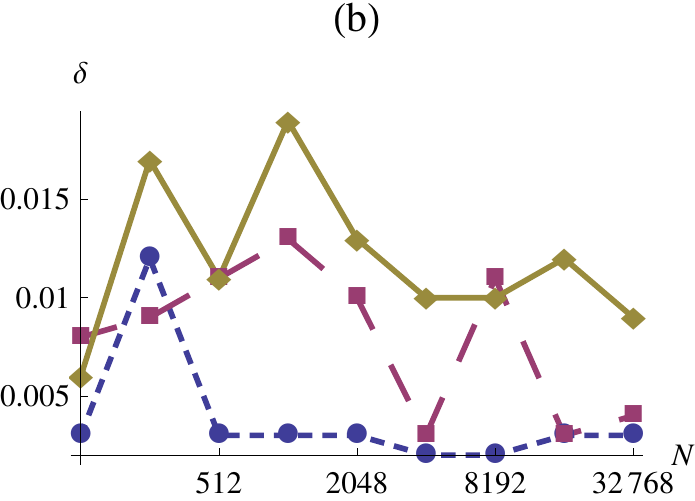}
\caption{\label{fig:results}The fraction ($\delta$) of times all paths between Alice and Bob were compromised, as a function of the number of nodes in the graph, for simulations using (a) uniform random graphs and (b) truncated power-law graphs. The yellow solid line, red long-dashed line, and blue short-dashed line represent values for $t=0.4$, $t=0.6$, and $t=0.8$, respectively.
}
\end{figure*}

One resource we have not yet explicitly discussed is bandwidth used during execution of the protocol. If we assume that the average number of times a given user initiates the protocol (i.e., acts as Alice) is fixed, it is clear that this will scale with the number of users involved in one instance of the protocol, which is given by the product of the path length with the number of paths. Using the numbers from Fig.~\ref{fig:parameters} and Table~\ref{table:longestpath} as an example, 
we see that, if $N=32768$ and $t=0.4$, a typical user would have to act as an intermediate node about $300$ times for every time he or she acted as a protocol initiator.
As only short keys ($\sim 1$ kb) are required for authentication, and the protocol need only be used the first time two mutual strangers wish to communicate, this does not represent an unreasonable burden on users.

Another cost borne by users is the necessity of establishing $\sim\!\!\langle k \rangle$ edges (shared secret keys) with other parties in order to join the network and be able to participate securely. For security reasons, the key exchanges required would have to take place offline via some secure means. Fortunately, the number of ``edges'' required is small, scaling as $O(\log N)$, so we believe this one-time cost will not be too onerous. New users could be advised of a minimum connectivity requirement and required to perform a certain number of secure offline key exchanges before being allowed to use the network.

\subsection{Network Topologies}\label{sec:topologies}

We have so far only considered a uniform random topology when discussing the application of the protocol to a network. However, in real-world networks, power-law graphs are common~\cite{albert:randomgraph}. Here we consider security in these ubiquitous, naturally-occurring network topologies.

Power-law graphs are those for which the number of nodes of degree $k$ is proportional to $k^{-\gamma}$, where $\gamma$ is a positive constant. When compared to a uniform random graph with the same number of nodes and edges, power-law graphs have been shown to be more robust against random removal of nodes~\cite{albert:randomgraph}; specifically, the size of the largest connected cluster decreases more slowly and the average path length in this cluster remains smaller (conversely, power-law graphs are more vulnerable than uniform random graphs to targeted node removal).
As Alice and Bob must still be connected within the subgraph induced by~$V(G)\backslash V_d$ in order to communicate securely using our protocol, this feature of power-law graphs suggests they are more secure against our adversary.

We modify the power-law structure slightly by imposing a minimum degree cut-off, $k_{\mathrm{min}}$; in practice, such a cut-off could be enforced by requiring all nodes to have at least $k_{\mathrm{min}}$ connections before allowing them to participate in the network. The cut-off is necessary because nodes with low degree are prone to protocol failure by compromise of all immediate neighbours (and, in a power-law graph, nodes of the lowest allowed degree are the most common); such failures are suppressed in a truncated power-law graph.

For example, given a node with degree $k$, the probability that all its neighbours are compromised (nc = neighbours compromised) is
\begin{equation}
\label{eq:P_nc}
	P_{\rm nc}(k)=(1-t)^k.
\end{equation}
Let us consider this particular protocol failure mode, in which all of Alice's immediate neighbours are compromised. We denote the probability of such failure by $\delta_{\rm nc}$. We then see that 
\begin{equation}
\delta_{\rm nc} = P_{\rm nc}(k_{\rm Alice}) \leq P_{\rm nc} (k_{\rm min}) = (1-t)^{k_{\rm min}}, \label{eq:mindeg}
\end{equation}
where $k_{\rm Alice}$ is Alice's degree and $k_{\rm min}$ is the minimum degree of any node in the network. It should be noted that $\delta_{\rm nc}$ represents only a portion of the total probability that the protocol will fail, $\delta$; by necessity, $\delta \geq \delta_{\rm nc}$. In the following section, we use Eq.~\req{eq:mindeg} to estimate the minimum degree cut-off $k_{\rm min}$ that should be imposed to constrain the probability of such failures to less than some allowed maximum $\delta_{\rm nc}$.

\section{Simulation results}

We now discuss the parameters and results of a simulation of the stranger authentication protocol. For each set of parameters, 1000 trials were run, with a new graph and new random choice of (non-compromised and non-adjacent) Alice and Bob for each trial. 

The number of paths was chosen according to Eq.~\req{eq:preq} and the number of edges according to Eq.~\req{eq:edges}, with $\delta=0.01$ used as the target probability of compromise. The variation of these parameters with changes in $t$ and $N$ is shown in Fig.~\ref{fig:parameters}. This procedure was followed for both uniform random and truncated power-law graphs, with the only difference being the type of graph; this allows us to perform a fair comparison of protocol performance on the two network topologies.

To generate uniform random graphs, pairs of nodes were randomly joined (with no duplicate edges allowed) until our quota of edges was met. To generate truncated power-law graphs, suitable minimum degrees were chosen according to Eq.~\req{eq:mindeg} with $\delta_{\rm nc} = 0.002$. The resulting values of $k_\mathrm{min}$ were 4, 7 and 12 for $t$ values of 0.8, 0.6 and 0.4, respectively. We chose the largest possible $\gamma$ such that the resulting number of edges in the graph, given by 
\begin{eqnarray}
\sum_{k=k_\mathrm{min}}^{k=N-1} \frac{NCk^{1-\gamma}}{2} \label{eq:plawedges}
\end{eqnarray}
was equal to or smaller than the goal number from Eq.~\req{eq:edges}. Degrees were assigned to each node such that the degree distribution of the graph was as close to $k^{-\gamma}$ as possible. Nodes were then randomly joined until all nodes satisfied their assigned degree as closely as possible (note that nodes were not allowed to exceed their assigned degree, which occasionally prevented a small number of nodes from reaching their assigned degree). The result was a truncated power-law graph with as many or slightly fewer edges than the corresponding uniform random graph for each set of parameters. 
We illustrate the difference between the degree distribution of the truncated power-law and uniform random topologies, generated according to the methods described above, in Fig.~\ref{fig:degreecomparison}.

Our simulation results are given in Table~\ref{table:longestpath} and Fig.~\ref{fig:results}. Table~\ref{table:longestpath} shows the logarithmic scaling of path length in $N$ suggested by Eq.~\req{eq:d}. This indicates that the cost of participating in the network, as outlined in Subsection \ref{sec:efficandcost}, scales reasonably (i.e., sub-exponentially). We see from Fig.~\ref{fig:results} that the probability of compromise is approximately constant in the uniformly random case; our numerical simulations match our predicted results in this regard. Probability of compromise does slowly increase with decreasing $t$ for both topologies. This behaviour is inevitable, as $\delta$ necessarily goes to 1 as $t$ goes to 0. We further see that truncated power-law graphs consistently offered greater security than uniform random networks for the adversarial model described in Section \ref{sec:adv_model}. This also confirms our expectations; truncated power-law graphs were shown to be more robust against random removal of nodes. In the alternate adversarial model in which the adversary is allowed to choose the compromised nodes, we expect power-law graphs to be somewhat less robust, due to the relative importance of a small number of highly connected nodes.

\section{Conclusion\label{sec:conclusion}}

We have shown a practical method for solving the stranger authentication problem, which arises when one attempts to build a large-scale secure communication network using QKD or cryptographic technology with similar properties. This method employs secret sharing to make use of multiple, partially-trusted paths through a network. Whereas the reliance on the collective integrity of many parties may at first seem surprising, it is preferable to current public key cryptographic strategies of using a root certificate authority who must always be trusted: our scheme avoids this single point of failure by not requiring any party, not even an authority, to be completely trusted. Through the choice of sufficiently many edges (secure authenticated channels) and paths, one can make the possibility of compromise vanishingly small.

Using both numerical simulations and random graph theory, we have shown that our protocol---in conjunction with QKD---provides security at reasonable resource costs. Protocol performance is shown via numerical simulations to be better in truncated power-law networks than in uniform random networks; since real communication and social networks tend to have a power-law structure, this bodes well for performance in the real world. Given that QKD systems are already commercially available, our methods could be implemented today. 

%\section{Acknowledgments}
\ack
We wish to thank L.\ Salvail, D.\ Stebila, and A.\ Roy for valuable discussions.
TRB acknowledges support from a US Department of Defense NDSEG Fellowship.
KPH appreciates financial support from the NSERC Undergraduate Student Research Award program.
BCS is a CIFAR Associate and acknowledges financial support from iCORE and NSERC.

%\section*{References}
\bibliographystyle{iopart-num}
\bibliography{authentication}

\end{document}